\documentclass{article}
\usepackage{waspaa19,amsmath,graphicx,url,times}
\usepackage{color}
\usepackage{graphics}

\usepackage{amsfonts}
\usepackage{here}
\usepackage{amssymb}
\usepackage{enumerate}
\usepackage{graphicx}
\usepackage{multirow}
\usepackage{tabularx}
\usepackage{arydshln}
\usepackage{color}
\usepackage{float}
\usepackage{bm}
\usepackage{comment}
\usepackage{cite}
\usepackage{amsmath}
\usepackage{algorithmic}
\usepackage{array}

\usepackage{setspace}

\newcommand{\bi}[1]{\ensuremath{\boldsymbol{#1}}}   

\newlength\savedwidth
\newcommand{\wcline}[1]{\noalign{\global\savedwidth\arrayrulewidth\global\arrayrulewidth 1.0pt} \cline{#1}
\noalign{\global\arrayrulewidth\savedwidth}}

\title{Joint Analysis of Acoustic Events and Scenes\\Based on Multitask Learning}

\name{Noriyuki Tonami$^{\dagger}$, Keisuke Imoto$^{\dagger}$, Masahiro Niitsuma$^{\dagger}$, Ryosuke Yamanishi$^{\dagger}$, and Yoichi Yamashita$^{\dagger}$}
\address{$^{\dagger}$ Ritsumeikan University, Shiga, Japan       
}

\begin{document}

\ninept
\maketitle


\begin{sloppy}

\begin{abstract}
Acoustic event detection and scene classification are major research tasks in environmental sound analysis, and many methods based on neural networks have been proposed.
Conventional methods have addressed these tasks separately; however, acoustic events and scenes are closely related to each other.
For example, in the acoustic scene ``office'', the acoustic events ``mouse clicking'' and ``keyboard typing'' are likely to occur.
In this paper, we propose multitask learning for joint analysis of acoustic events and scenes, which shares the parts of the networks holding information on acoustic events and scenes in common. 
By integrating the two networks, we expect that information on acoustic scenes will improve the performance of acoustic event detection.
Experimental results obtained using TUT Sound Events 2016/2017 and TUT Acoustic Scenes 2016 datasets indicate that the proposed method improves the performance of acoustic event detection by 10.66 percentage points in terms of the F-score, compared with a conventional method based on a convolutional recurrent neural network.
\end{abstract}
%
%
\begin{keywords}
Acoustic event detection, acoustic scene classification, convolutional recurrent neural network, multitask learning
\end{keywords}
%
%
\section{Introduction}
\label{sec:intro}
There has been increasing interest in analyzing various sounds in real-life environments such as cooking sounds in a kitchen or car passing sounds in a street \cite{Imoto_AST2018_01}.
Various applications can be expected from the automatic analysis of environmental sounds, for example, abnormal sound detection systems \cite{abnormal}, automatic life-logging systems \cite{Stork_ROMAN2012_01,Imoto_INTERSPEECH2013_01}, surveillance systems \cite{surveillance,Ntalampiras_ICASSP2009_01}, and hearing-impaired support systems \cite{healthcare,water}.
%
In environmental sound analysis, the following two tasks have mainly been studied: acoustic event detection (AED) and acoustic scene classification (ASC).
AED is the task of detecting acoustic event labels and their onset/offset in an audio recording, where an acoustic event indicates a type of sound such as ``mouse clicking,'' ``people talking,'' or ``bird singing.''
ASC is the task of predicting acoustic scene labels from a relatively long duration recording, where an acoustic scene indicates a recording situation with human activity such as ``office,'' ``train,'' or ``cooking.''

For AED and ASC, many methods based on the Gaussian mixture model (GMM) \cite{AED_GMM,ASC_GMM}, hidden Markov model (HMM) \cite{AED_HMM,ASC_HMM}, and support vector machine (SVM) \cite{ASC_SVM}, have been proposed.
As alternative approaches for AED, those based on non-negative matrix factorization (NMF) \cite{AED_NMF,Komatsu_DCASE2016_01} have been studied.
Such approaches enable the analysis of polyphonic acoustic events by decomposing their spectrum into product of a basis and activation matrix, where each basis and each activation vector indicate a single acoustic event and the active duration of the corresponding event, respectively.
More recently, developments in acoustic event and scene analysis have led to renewed interest in deep neural networks (DNNs) \cite{DCASE2017_AED,DCASE2017_ASC}.
The DNN-based approaches achieve better results than the conventional methods based on HMM, GMM, and SVM.

In AED and ASC, most of the studies have addressed acoustic events and scenes analysis separately; in many cases however, acoustic events and scenes are related to each other.
For example, in the acoustic scene ``office'', the acoustic events ``mouse clicking'' and ``keyboard typing'' are likely to occur, whereas the acoustic events ``car'' and ``bird singing'' do not tend to occur.
%
In other words, when analyzing the acoustic events ``mouse clicking'' and ``keyboard typing,''  it is expected that information on the acoustic scene ``office'' will help in detecting these acoustic events.
%
On the basis of this idea, AED utilizing information of the acoustic scene in an unsupervised manner \cite{Mesaros_EUSIPCO2011_01,Imoto_IEICE2016_01} and ASC taking information of acoustic events into account \cite{Imoto_TASLP2019_01}, which are based on Bayesian generative models, have been proposed.
However, these conventional methods do not estimate both acoustic events and scenes  explicitly.
Moreover, these methods cannot be applied to state-of-the-art neural network-based methods.

In this paper, we present a new method for the joint analysis of acoustic events and scenes based on multitask learning.
The contribution of this work is summarized as follows:
\begin{itemize}
  \item we propose a method for joint analysis of acoustic events and scenes using multitask learning combining state-of-the-art AED and ASC approaches.
  \item we demonstrate that the multitask learning-based method improves the AED performance.
\end{itemize}

The remainder of this paper is structured as follows.
In Sec. 2, conventional methods for AED and ASC are first discussed and the proposed method for joint analysis of acoustic events and scenes based on multitask learning is then introduced.
In Sec. 3, the experiments carried out to evaluate the performance of event detection and scene classification are reported.
Finally, we summarize and conclude this paper in Sec. 4.
%
%
\section{Multitask Learning of Acoustic Events and Scenes}
\subsection{Conventional Method for Event Detection and Scene Classification}
\label{ssec:conventional}
In this section, we review conventional AED and ASC methods based on neural networks.
AED involves the estimation of acoustic event labels and their onset/offset times, where acoustic events may overlap in the time axis.
In recent years, many neural network based methods, such as a convolutional neural network (CNN)-based approach \cite{Hershey_ICASSP2017_01} and a recurrent neural network (RNN)-based approach \cite{Hayashi_TASLP2017_01}, have been proposed.
Specifically, it has been reported that neural networks combining a CNN and a bidirectional gated recurrent unit (BiGRU) successfully analyze acoustic events with reasonable performance \cite{SED_CRNN,Imoto_ICASSP2019_01}.
In the CNN-BiGRU network, the acoustic feature ${\bf X} \in \mathbb{R}^{D \times T}$, which is a time and frequency--spectrum representation, is input to the network.
Here, $D$ and $T$ are the number of frequency bins and the number of time frames of the input acoustic feature, respectively.
The convolution layer convolutes the input feature map with two-dimensional filters, then max pooling is conducted to reduce the dimension of the feature map.
The output of the convolutional layer ${\bf V} \in \mathbb{R}^{D' \times T \times C}$ is then concatenated as ${\bf V'} = ({\bf x}_{1}, {\bf x}_{2}, \ldots{\bf x}_{t}, \ldots, {\bf x}_{T}) \in \mathbb{R}^{(D' \cdot C)\times T}$, and then ${\bf V'}$ is fed to the BiGRU layer, where $t$ and $C$ are respectively the time index and the number of filters of the convolution layer.
After that, the output vector ${\bf h}_{t}$ is calculated using the following equations:

%
\begin{align}
{\bf g}^{f}_{t} &= \sigma({\bf W}^{f}_{g} {\bf x}_{t} + {\bf U}^{f}_{g} {\bf h}_{t-1} + {\bf b}^{f}_{g}),\\[1pt]
{\bf r}^{f}_{t} &= \sigma({\bf W}^{f}_{r} {\bf x}_{t} + {\bf U}^{f}_{r} {\bf h}_{t-1} + {\bf b}^{f}_{r}),\\[1pt]
{\bf h}^{f}_{t} &= (1-{\bf g}^{f}_{t}) \odot {\bf h}_{t-1} \nonumber\\[-2pt]
&\hspace{10pt}+ {\bf g}^{f}_{t} \odot \tanh ({\bf W}^{f}_{h} {\bf x}_{t} + {\bf U}^{f}_{h} ( {\bf r}^{f}_{h} {\bf h}_{t-1}) + {\bf b}^{f}_{h}),\\[1pt]
{\bf g}^{b}_{t} &= \sigma({\bf W}^{b}_{g} {\bf x}_{t} + {\bf U}^{b}_{g} {\bf h}_{t+1} + {\bf b}^{b}_{g}),\\[1pt]
{\bf r}^{b}_{t} &= \sigma({\bf W}^{b}_{r} {\bf x}_{t} + {\bf U}^{b}_{r} {\bf h}_{t+1} + {\bf b}^{b}_{r}),\\[1pt]
{\bf h}^{b}_{t} &= (1-{\bf g}^{b}_{t}) \odot {\bf h}_{t+1} \nonumber\\[-2pt]
&\hspace{10pt}+ {\bf g}^{b}_{t} \odot \tanh ({\bf W}^{b}_{h} {\bf x}_{t} + {\bf U}^{b}_{h} ( {\bf r}^{b}_{h} {\bf h}_{t+1}) + {\bf b}^{b}_{h}),\\[1pt]
{\bf h}_{t} &=
\begin{bmatrix}
{\bf h}^{f}_{t}\\[3pt]
{\bf h}^{b}_{t}
\end{bmatrix},
\label{eq:bigru}
\end{align}
\vspace{3pt}

\noindent where ${\bf W}$ and ${\bf U}$ are parameter matrices and ${\bf b}$ is a bias vector.
Superscripts $f$ and $b$ are the forward and backward networks, respectively.
Subscripts $g$ and $r$ indicate the update gate and reset gate, respectively.
${\bf g}$, ${\bf r}$, $\odot$, and $\sigma$ indicate the update gate vector, reset gate vector, Hadamard product, and sigmoid function, respectively.
The BiGRU layer is followed by a fully connected layer, which is the output layer of the network calculated as

%
\begin{align}
{\bf y}_{t} &= \sigma({\bf h}_{t}).
\label{eq:outputlayer_event}
\end{align}
%

\noindent The CNN-BiGRU network is optimized under the following sigmoid cross-entropy objective function $E_{1}({\bi \Theta}_{1})$ using the backpropagation through time (BPTT):

%
\begin{align}
E_{1}({\bi \Theta}_{1}) &= - \! \sum^{T}_{t=1} {\big \{} {\bf z}_{t} \log ( {\bf y}_{t} ) + (1-{\bf z}_{t}) \log (1-{\bf y}_{t}) {\big \}} \nonumber\\[-1pt]
&= - \! \sum^{M}_{m=1} \sum^{T}_{t=1} \! {\Big \{} z_{m,t} \log y_{m,t} \! + \! (1 \! - \! z_{m,t}) \log {\big (} 1 \! - \! y_{m,t} {\big )} \! {\Big \}},
\label{eq:event_loss}
\end{align}

\noindent where $m$ and $z_{m,t}$ are the index of the acoustic event category and the target label in time frame $t$, which is 1 if acoustic event $m$ is active in time frame $t$, and 0 otherwise.
%
%
%
%
%
%
%

ASC is the task of estimating the acoustic scene with which a sound clip is most associated.
It has been reported that CNN-based methods achieve state-of-the-art performance in ASC \cite{Mesaros2017,Sakashita_DCASE2018_01}
%
%
%
In CNN-based ASC, the time-frequency representation of acoustic feature ${\bf X}$ is fed to a convolutional layer.
The CNN layer is followed by a fully connected layer, which is the output layer of the network calculated as

%
\begin{align}
{\bf y}_{t} &= S({\bf h}_{t}),
\label{eq:outputlayer_scene}
\end{align}
%

\noindent where $S$ indicates the softmax function.
This network is optimized under the following softmax cross-entropy objective function $E_{2}({\bi \Theta}_{2})$:

%
\begin{align}
E_{2}({\bi \Theta}_{2}) = - \sum^{N}_{n=1} {\Big \{} z_{n} \log S (y_{n}) {\Big \}},
\label{eq:scene_loss}
\end{align}

\noindent where $n$ and $z_n$ are the number of acoustic scene categories and the target label, respectively.
%
%
\begin{figure}[t!]
\centering
\includegraphics[width=0.98\columnwidth]{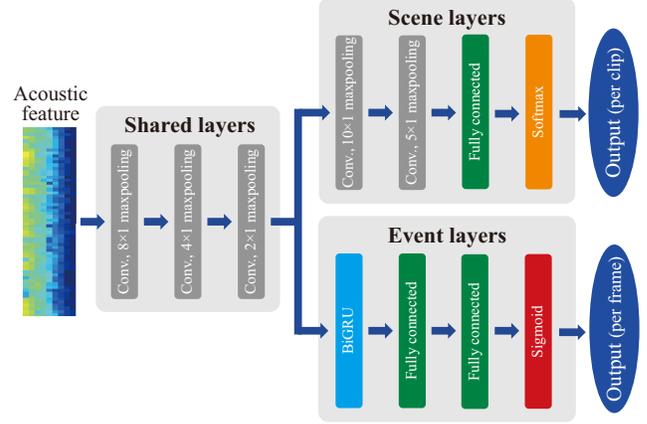}
\vspace{-2pt}
\caption{Example of network structure for proposed multitask learning of acoustic events and scenes}
\label{fig:network}
\end{figure}
%
%
%
\subsection{Proposed Method}
%
%
The conventional methods for AED and ASC have been studied separately.
However, many acoustic events and scenes are related; thus, we consider that the middle layers of the networks for both AED and ASC extract features common to the acoustic event and scene.
On the basis of this idea, we propose the multitask learning of acoustic events and scenes, which shares the parts of the networks holding information of acoustic events and scenes in common.
The multitask learning enables us to jointly analyze multiple tasks that are related to each other \cite{Caruana1997}.
In speech processing, it has been reported that multitask learning successfully improves the performance of speech recognition \cite{Giri_ICASSP2015_01,Kim_ICASSP2017_01}.
Moreover, it has also been reported that when there is a difference in the degree of difficulty between tasks, the performance of the harder task is improved using the information obtained from the easier task \cite{Caruana1997}.
In this work, we consider that AED is the harder task and ASC is the easier task, and expect that the information of acoustic scenes will help AED.
%
%

The concept of the proposed method is shown in Fig.~\ref{fig:network}, where we refer to the shared part of the network as ``shared layers.''
In this network, we apply the CRNN as the event detection network and the CNN as the scene classification network, and the CNN layers are shared between the event detection and scene classification networks.

The objective function for the proposed method is expressed as

\begin{align}
E({\bi \Theta}) = E_{1}({\bi \Theta}_{1}) + \alpha E_{2}({\bi \Theta}_{2}),
\label{eq:loss}
\end{align}

\noindent where $\alpha$ is the weight of the acoustic scene classification loss.
Note that the proposed method can also be applied to different networks from that shown in Fig.~\ref{fig:network} when the networks can be optimized by Eq.~(\ref{eq:loss}).
%
%
%
\section{Evaluation Experiments}
\vspace{-2pt}
\subsection{Experimental Conditions}
We evaluated the performance of AED and ASC using the TUT Sound Events 2016 \cite{Mesaros2016TUTDF} development, the 2017 \cite{Mesaros2017} development, and the TUT Acoustic Scenes 2016 \cite{Mesaros2016TUTDF} development datasets.
In these datasets, we used sound clips including four acoustic scenes; ``home,'' ``residential area'' (TUT Sound Events 2016), ``city center'' (TUT Sound Events 2017), and ``office'' (TUT Acoustic Scenes 2016), which contain 192 min of audio.
Here, the acoustic scene ``office'' did not have acoustic event labels; thus, we manually annotated event labels to the sound clips with the protocol described in \cite{Mesaros2016TUTDF} and \cite{Mesaros2017}.
These sound clips contain the 25 acoustic events listed in Table~\ref{tbl:each_event}.

We used log-mel energies of 64 dimensions as acoustic features, which were calculated for each 40 ms time frame with 50\% overlap.
The acoustic features were then input to the network shown in Fig.~\ref{fig:network}.
AED and ASC results were obtained from the outputs of the network, in which active acoustic events were predicted using 0.5 as the threshold of activation.
The experiments were conducted using a four-fold cross-validation setup.
Other experimental conditions are listed in Table~\ref{tbl:parameter}.  

\begin{table}[t]
\small
\caption{Experimental conditions}
\vspace{-15pt}
\label{tbl:parameter}
\begin{center}
\begin{tabular}{ll}
\wcline{1-2}
&\\[-10pt]
Acoustic feature & Log-mel energy (64 dim.)\\
Frame length \hspace{-3pt} / \hspace{-3pt} shift & 40 ms \hspace{-3pt} / \hspace{-3pt} 20 ms\\
Length of sound clip & 10 s\\\hline
&\\[-10pt]
Network structure of shared layers & 3 CNN \\
\# channels of CNN layers (shared) & 128, 128, 128 \\
Filter size (shared) & 1$\times$3 \\
Pooling size (shared) & 8, 4, 2 (max pooling) \\\hline
&\\[-10pt]
Network structure of scene layers & 2 CNN \\
\# channels of CNN layers (scene) & 64, 16 \\
Filter size (scene) & 3$\times$3 \\
Pooling size (scene) & 10, 5 (max pooling) \\\hline
&\\[-10pt]
Network structure of event layers & 1 BiGRU \& 1 fully conn. \\
\# units in GRU layer (event) & 32 \\
\# units in fully conn. layer (event) & 32 \\
\wcline{1-2}
\end{tabular}
\vspace{-15pt}
\end{center}
\end{table}
%

\subsection{Metrics}
In AED, since acoustic events may overlap, the event detection performance is evaluated on the basis of a segment-based F-score and error rate (ER) \cite{Mesaros2016_MDPI}.
To calculate the segment-based F-score, the precision and recall are first calculated as

\vspace{-2pt}
\begin{align}
\label{eq:Precision}
{\rm Precision} =  \frac{{\rm TP}}{{\rm TP}+{\rm FP}},
\vspace{-5pt}
\end{align}
\begin{align}
{\rm Recall} = \frac{{\rm TP}}{{\rm TP}+{\rm FN}},
\label{eq:Recall}
\end{align}

\noindent where TP, FP, and FN are the total counts of true positive, false positive, and false negative for all time frames and acoustic events, respectively.
The segment-based F-score is then calculated as 

\begin{align}
{\rm F\textrm{-}score} = \frac{2 \cdot {\rm Precision} \cdot {\rm Recall}}{{\rm Precision} + {\rm Recall}}.
\label{eq:F}
\end{align}

%
%
\begin{table}[t]
\small
\caption{Performance of acoustic event detection and scene classification}
\vspace{-4pt}
\label{tbl:F-score}
\begin{center}
\begin{tabular}{cccc}
\wcline{1-4}
&&&\\[-10pt]
\multirow{2}{*}{Method} & \multicolumn{2}{c}{Event} & Scene \\\cline{2-4}
&&&\\[-10pt]
& F-score & ER  & F-score \\\hline
&&&\\[-10pt]
\multicolumn{1}{l}{CRNN (event)} & 38.90\% & 0.776 & -\\
\multicolumn{1}{l}{CNN (scene)} & - & - & {\bf 66.36\%} \\\hline
&&&\\[-10pt]
\multicolumn{1}{l}{Multitask ($\alpha$=0.1)}  & 44.31\% & 0.721 & 46.72\% \\
\multicolumn{1}{l}{Multitask ($\alpha$=0.01)} & {\bf 49.56\%} & {\bf 0.695} & 58.94\% \\
\multicolumn{1}{l}{Multitask ($\alpha$=0.001)} & 41.11\% & 0.760 & 52.93\% \\ \wcline{1-4}
\end{tabular}
\vspace{-5pt}   
\end{center}
\end{table}
%
%
\begin{figure}[t!]
\centering
\includegraphics[width=1.00\columnwidth]{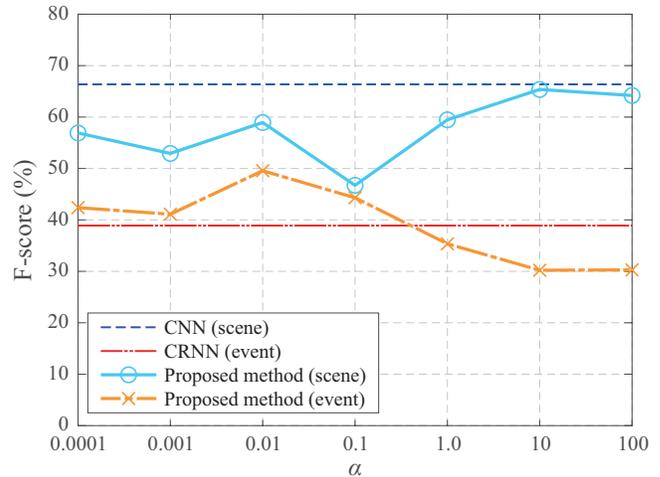}
\vspace{-18pt}
\caption{Acoustic event detection and scene classification performance as functions of weight $\alpha$}
\label{fig:alpha_relation}
\end{figure}
%
%
%
\begin{table*}[t]
\small
\caption{Acoustic event detection performance for each event}
\vspace{3pt}   
\centering
\scalebox{0.8}[0.8]{
\begin{tabular}{llcccccccccccc}
\wcline{1-14}
&&&&&&&&&&&&&\\[-8pt]
\multicolumn{2}{c}{\multirow{2}{*}{Event}} & (object) & (object) & (object) & (object) & (object) & bird & brakes & \multirow{2}{*}{breathing} & \multirow{2}{*}{car} & \multirow{2}{*}{children} & \multirow{2}{*}{cupboard} & \multirow{2}{*}{cutlery}
\\[-1pt]
&& banging & impact & rustling & snapping &  squeaking & singing & squeaking &&&&&\\
&&&&&&&&&&&&&\\[-20pt]
\\\hline
&&&&&&&&&&&&&\\[-8pt]
CRNN & F-score & 0.00\% & 3.83\% & 0.00\% & 0.00\% & 0.00\% & 51.84\% & 17.86\% & 0.00\% & 67.24\% & 0.00\% & 0.00\% & 0.00\% \\\cline{2-14}
(event)&&&&&&&&&&&&&\\[-8pt]
 & ER ($\times 10^{-3}$) & 1.05 & 26.12 & 24.67 & 1.76 & 1.52 & 89.32 & 10.19 & 1.93 & 132.87 & 25.49 & 1.66 & 3.96\\\hline
\\[-8pt]
Proposed& F-score & 0.00\% & 4.01\% & 0.00\% & 0.00\% & 0.00\% & 52.52\% & 0.83\% & 0.00\% & 63.05\% & 0.00\% & 0.00\% & 0.00\% \\\cline{2-14}
($\alpha$=0.01)&&&&&&&&&&&&&\\[-8pt]
 & ER ($\times 10^{-3}$) & 1.05 & 26.06 & 24.67 & 1.76 & 1.52 & 91.12 & 10.57 & 1.93 & 138.76 & 25.49 & 1.66 & 3.96\\
\wcline{1-14}
\end{tabular}
}
%
\vspace{8pt}
%
\small
\label{tbl:each_event}
\centering
\scalebox{0.777}[0.777]{
\begin{tabular}{llccccccccccccc}
\wcline{1-15}
&&&&&&&&&&&&&&\\[-8pt]
\multicolumn{2}{c}{\multirow{2}{*}{Event}} & \multirow{2}{*}{dishes} & \multirow{2}{*}{drawer} & \multirow{2}{*}{fan} & glass & keyboard & large & mouse & mouse & people & people & washing & water tap & wind
\\[-1pt]
&&&&& jinging & typing & vehicle & clicking & wheeling & talking & walking & dishes & running & blowing\\
&&&&&&&&&&&&&&\\[-20pt]
\\\hline
&&&&&&&&&&&&&&\\[-8pt]
CRNN & F-score & 15.99\% & 0.00\% & 60.45\% & 0.00\% & 4.05\% & 51.45\% & 0.00\% & 0.00\% & 0.03\% & 42.68\% & 34.65\% & 43.48\% & 0.00\% \\\cline{2-15}
(event)&&&&&&&&&&&&&&\\[-8pt]
 & ER ($\times 10^{-3}$) & 11.04 & 1.76 & 159.77 & 1.98 & 24.53 & 56.07 & 6.89 & 2.35 & 98.82 & 119.87 & 21.93 & 16.77 & 13.42 \\\hline
\\[-8pt]
Proposed & F-score & 20.70\% & 0.00\% & 64.31\% & 0.00\% & 8.21\% & 53.50\% & 1.45\% & 0.00\% & 0.00\% & 49.97\% & 45.40\% & 40.68\% & 2.86\% \\\cline{2-15}
($\alpha$=0.01)&&&&&&&&&&&&&&\\[-8pt]
 & ER ($\times 10^{-3}$) & 10.69 & 1.76 & 165.03 & 1.98 & 24.30 & 52.97 & 6.87 & 2.35 & 98.82 & 105.14 & 21.15 & 16.98 & 13.30 \\
\wcline{1-15}
\end{tabular}
}
\vspace{3pt}
\end{table*}

\noindent To calculate the segment-based ER, substitutions (S), deletions (D), and insertions (I) are first calculated as

\begin{align}
{\rm S}(k) &= \min({\rm FN}(k), {\rm FP}(k)),\\
{\rm D}(k) &= \max(0, {\rm FN}(k)-{\rm FP}(k)),\\
{\rm I}(k) &= \max(0, {\rm FP}(k)-{\rm FN}(k)),
\label{eq:SDI}
\end{align}

\noindent where $k$ indicates the index of the time frame.
The ER is then calculated as

\begin{align}
{\rm ER} =  \frac{\sum_{k=1}^{K} {\rm S}(k) + \sum_{k=1}^{K} {\rm D}(k) + \sum_{k=1}^{K} {\rm I}(k)}{\sum_{k=1}^{K} {\rm N}(k)},
\label{eq:ER}
\end{align}

\noindent where $K$ and $N(k)$ are the total number of time frames and the number of acoustic events in time frame $k$, respectively.

On the other hand, in ASC, acoustic scenes do not overlap; thus, it is regarded as a simple classification task.
The F-score is calculated using Eqs.~(\ref{eq:Precision}) -- (\ref{eq:F}), where TP, FP, and FN are the total counts of true positive, false positive, and false negative for all sound clips, respectively.
%
%
\subsection{Experimental Results}
%
As comparative methods, we evaluated the detection of an event using CRNN (referred to as CRNN (event)) and the classification of a scene using CNN (referred to as CNN (scene)).
CRNN (event) had the same structures and parameters as those of the shared and event layers in Fig.~\ref{fig:network}, whereas CNN (scene) had the same structures and parameters as those of the shared and scene layers.

Experimental results are shown in Table~\ref{tbl:F-score}.
The results show that the proposed multitask-based method enables the joint analysis of acoustic events and scenes with a reasonable performance compared with CRNN (event) and CNN (scene).
In particular, when $\alpha$ = 0.01, the proposed method improved the performance by 10.66 percentage points in terms of the F-score of the event detection result compared with that of the conventional method.
This result indicates that information of acoustic scenes improves the performance of AED.

Further evaluations were conducted to investigate how the proposed method performs with various settings of $\alpha$.
The performances in event detection and scene classification for various $\alpha$ are shown in Fig.~\ref{fig:alpha_relation}.
The results show that when $\alpha$ is less than 1.0, the performance of AED is better than  that of the conventional CRNN (event).
When $\alpha$ is larger than 0.1, the performance of ASC improved, whereas that of AED was worse than that of the conventional CRNN (event).
This implies that AED is a harder task than ASC, and thus, the parameter $\alpha$ should be set to a small value to achieve AED with better performance than the conventional methods.

For a more detailed examination, we list the detection results for each event in Table~\ref{tbl:each_event}.
In this experiment, we evaluated the detection performance using $\alpha$ = 0.01.
The results show that the proposed method improves the F-score and ER in many of the acoustic events.
For example, the acoustic events ``dishes'', ``people walking,'' and ``washing dishes'' can be detected more accurately by the proposed method; the F-scores of these acoustic events increase by 4.71, 7.29, and 10.75 percentage points, respectively, compared with the conventional CRNN (event).
On the other hand, the detection performance for ``brakes squeaking'' and ``car'' do not improve.
This is because they may occur in both the acoustic scenes ``residential area'' and ``city center,'' and thus, information of acoustic scenes may not effectively improve the performance of AED.
%
%
%
%
\section{Conclusion}
In this paper, we proposed multitask learning for the joint analysis of acoustic events and scenes.
In the proposed method, we applied a CRNN as the event detection network and a CNN as the scene classification network, and the CNN layers in both network were shared.
Then, we integrated their objective functions using a weight parameter and optimized the network simultaneously.
Experimental results indicated that the proposed method enables joint analysis of acoustic events and scenes with reasonable performance compared with the conventional methods.
Moreover, the proposed method improves the performance of acoustic event detection by 10.66 percentage points in terms of the F-score and by 0.081 in terms of ER, compared with a conventional CRNN-based method.
%
\section{Acknowledgement}
\label{sec:ack}
This work was supported by JSPS KAKENHI Grant Number JP19K20304.
%
\small
\bibliographystyle{IEEEtran}
\bibliography{IEEEabrv,refs19}

\end{sloppy}
\end{document}